\documentclass{ifacconf}

\usepackage{graphicx}      
\usepackage{natbib}        

\usepackage{comment}
\usepackage{amsmath}
\usepackage{nccmath}
\usepackage{array}
\usepackage{color}
\usepackage{amssymb}
\usepackage{mathrsfs}
\usepackage{amsfonts, mathtools}
\usepackage{soul}
\usepackage{verbatim}
\usepackage{siunitx}
\usepackage{multirow}
\usepackage[table]{xcolor}
\usepackage[numbered,framed]{matlab-prettifier}
\usepackage{subcaption} 
\usepackage{float}
\usepackage{multicol}
\usepackage{lipsum}
\usepackage{csquotes}
\usepackage{arydshln}
\usepackage{xspace}
\usepackage[normalem]{ulem}

\let\rm\mathrm
\let\bf\mathbf
\let\sf\mathsf
\let\bb\mathbb
\let\cal\mathcal

\definecolor{mblue}{rgb}{0,0.4470,0.7410}
\definecolor{mgreen}{RGB}{22,171,53}

\newcommand{\normal}{\mathcal{N}}
\newcommand{\tss}[1]{\textsuperscript{#1}}
\newcommand{\mr}[1]{\mathrm{#1}}
\newcommand{\dnth}{n_\theta}

\newcommand{\chris}[1]{{#1}}
\newcommand{\todo}[1]{{#1}}
\newcommand{\matthis}[1]{{#1}}
\begin{document}
\begin{frontmatter}

\title{\hspace*{-29.11117pt}\mbox{LPV Modeling of the Atmospheric Flight Dynamics} of \matthis{a Generic Parafoil Return Vehicle} 
\thanksref{footnoteinfo}} 

\thanks[footnoteinfo]{This work was supported by the European Space Agency in the scope of the `AI4GNC' project with SENER Aeroespacial S.A. (contract nr. 4000133595/20/NL/CRS) and was also partly supported by Ministry of Innovation and Technology NRDI Office within the framework of the Autonomous Systems National Laboratory Program. The views expressed in this paper do not reflect the official opinion of the European Space Agency. Corresponding author: Matthis de Lange (\texttt{m.h.d.lange@student.tue.nl})}

\author[tue]{Matthis H. de Lange}
\author[tue]{Chris Verhoek}
\author[esa]{Valentin Preda} 
\author[tue,sztaki]{Roland T{\'o}th}

\address[tue]{Control Systems Group, Dept. of Electrical Engineering, Eindhoven University of Technology, Eindhoven 5600MB, The Netherlands.}
\address[esa]{\mbox{ESTEC, European Space Agency, Noordwijk 2200AG, The Netherlands.}}
\address[sztaki]{Systems and Control Lab, Institute for Computer Science and Control, Budapest 1111, Hungary.}

\begin{abstract}
%
Obtaining models that can be used for control is of utmost importance to ensure the guidance and navigation of spacecraft, like a \emph{Generic Parafoil Return Vehicle} (GPRV). In this paper, we convert a nonlinear model of the atmospheric flight dynamics of an GPRV to a \emph{Linear Parameter-Varying} (LPV) description, such that the LPV model is suitable for navigation control design. Automated conversion methods for nonlinear models can result in complex LPV representation, which are not suitable for controller synthesis. We apply several state-of-the-art techniques, including learning based approaches, to optimize the complexity and conservatism of the LPV embedding for an GPRV. The results show that we can obtain an LPV embedding that approximates the complex nonlinear dynamics sufficiently well, where the balance between complexity, conservatism and model performance is optimal.
\end{abstract}


\end{frontmatter}

\section{Introduction}

\begin{figure}
\centering
\includegraphics[width=\linewidth]{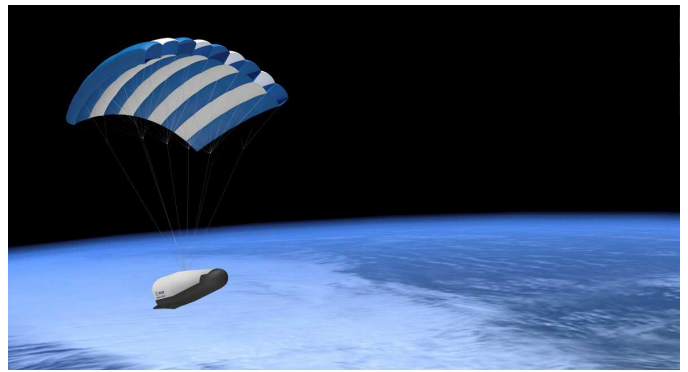}
\caption{Space Rider reusable spacecraft navigating with its parafoil during the final phase of the re-entry and landing process. (Image taken from \texttt{https://esa.int})} 
\label{fig:Intro:SR}
\end{figure}

The \emph{European Space Agency} (ESA) is currently developing a \emph{Generic Parafoil Return Vehicle} (GPRV) to perform missions at low orbits, which is designed to re-enter Earth's atmosphere and land at a designated location on the surface, such that the vehicle can be reused. An example of such a vehicle is the \emph{Space Rider} reusable spacecraft. In the final stage of the landing process, the \matthis{GPRV} is navigated towards the landing point by a guided parafoil as shown in Figure \ref{fig:Intro:SR}. The navigation is challenging, as the flight dynamics of the \matthis{GPRV} are subject to changing aerodynamical effects, the parafoil itself is attached by flexible tension lines to the canopy whose motion is governed by complex fluid dynamics, and the overall vehicle is subject to harsh wind disturbances, while the only available actuation is steering via the parafoil (there is no active propulsion). 

Reliable and accurate motion control of  is essential for proper navigation (\matthis{heading and} flight path tracking) and guaranteeing a safe landing. For this reason, the development of an accurate model, useful for flight controller design, is crucial for the \emph{Guidance, Navigation \& Control} (GNC) development of the prototype. In this research, we focus on obtaining a high-fidelity model of  for this purpose and show how it can be converted to \emph{Linear Parameter-Varying} (LPV) forms with various complexity levels. 

Common control design strategies in aerospace applications rely heavily on the LPV framework \cite[]{wu1995lpv, corti2012lpv}. This is due to the fact that with this framework it is possible to embed complex nonlinear systems in a representation with \emph{linear} signal relations. These relations however vary with a time-varying, measurable signal $p$, called the \emph{scheduling}. This allows to extend powerful methods of the \emph{Linear Time-Invariant} (LTI) framework to design controllers with stability and performance guarantees and rely on efficient performance shaping concepts. While linearity of the resulting LPV surrogate models of the dynamics enables simplified control design and efficient performance shaping, the construction of the often multidimensional $p$ that describes the effect of nonlinearities is highly important. In LPV embedding, the constructed scheduling is inherently dependent on internal latent variables, like states and inputs of the system, whose relationship is excluded from the LPV model, intentionally seeing it as an external independent variable \cite[]{Toth2010}. By assuming all possible variations of $p$, the solution set of the LPV model will include the original trajectories of the nonlinear model, but possibly even more due to the disregarded relationship between $p$ and its inducing variables. This is called \emph{conservativeness} of the embedding, and its reduction is highly important to avoid deterioration of the achievable performance of LPV control based on the extracted  surrogate model \cite[]{Toth2010}. Furthermore, the dimension of the constructed $p$ and functional dependence of the LPV model coefficients on $p$ (e.g., affine, polynomial, etc. dependence of the matrices in a state-space representation), i.e., \emph{complexity} of the LPV model, have major impact on the computability of model-based LPV controller synthesis \citep{hoffmann2014survey}. Hence, reduction of such complexity is also a key objective of the LPV modeling toolchain. For this purpose, several conversion strategies, e.g., \emph{substitution based transformation} (SBT) methods \citep{shamma2000gain_scheduling,Carter1996,Marcos2004} and \emph{automated conversion procedures} \citep{Kwiatkowski2006Automated_generation,Hoffmann2015LPV_Gyro,Toth2010} together with various complexity reduction methods, e.g., \citep{beck2006modelbasedred_lpv_coprime, hecker2005symbolic}, have been introduced and also applied for spacecraft models in \cite{varga1998modelbasedred_aircraft}. However, only a limited number of methods have been derived to optimize the scheduling complexity in the conversion process, like the family of \emph{Principle Component Analysis} (PCA) methods  \citep{Kwiatkowski2008, rizvi2016kernelpca, sadeghzadeh2020affine} and learning-based scheduling reduction methods discussed in \cite{rizvi2018autoencoder, Koelewijn2020}.

In \cite{DeLange2021}, the 12 \emph{Degree of Freedom} (DOF) motion dynamics of  with the parafoil, including 
detailed aerodynamical effects on the two bodies, have been derived in terms of a  nonlinear dynamical model. To make this model suitable for LPV control, as a main contribution of the paper, we develop a global embedding of these dynamics in terms of an LPV representation, where both the conservativeness and complexity of the embedding are optimized. For this purpose, we apply and compare two data-based scheduling dimension reduction methods: (i) the PCA method in \citep{sadeghzadeh2020affine} that is the current state-of-the-art method in the PCA family of reduction techniques and the (ii) \emph{Deep Neural Network} (DNN) method from \citep{Koelewijn2020} that has been reported to perform the best among the learning based methods. The accuracy of the obtained LPV models with various complexity levels is analyzed in simulation studies with the original model. 

The paper is organized as follows. First in Section \ref{sec:SR:mod}, the dynamical motion model of  is introduced and its LPV conversion based on a direct factorization approach is explained. This is followed in Section \ref{s:prelim} by a brief overview of the PCA method and the DNN method used for complexity and conservativeness reduction of the converted LPV model. In Section \ref{sec:sr_of_sr}, the discussed methods are applied on the LPV modeling problem of  and the results are analyzed together with simulation based verification of the resulting models w.r.t. the original high-fidelity model. Finally, the conclusions on obtained results are given in Section \ref{sec:conclusion}.

\section{Modeling of the Flight Dynamics}  \label{sec:SR:mod}
%
\subsection{High-fidelity dynamical model}\label{ss:highfidmodel}
The model of the flight dynamics of the  attached to the parafoil with flexible tension lines and a spherical joint is derived based on \cite{Yakimenko2015a}, and \cite{Figueroa-Gonzalez2021a} in \cite{DeLange2021} using variable air density and the specific aerodynamic properties of the \matthis{GPRV}. A simplified form of this model, where the parafoil and the body of the space craft is approximated as a single rigid body, corresponds to the dynamic motion model
\matthis{ 
\begin{subequations}
\label{eq:under:model}
\begin{align}
		\dot{r} &= R(\eta)v\\
		\dot{\eta} &= J(\eta)\omega\\
		\dot{v} &= -\omega\times v+\frac{1}{m}\left(f_\rm{a}(\eta,v,\omega,\delta,w)+f_\rm{g}(r,\eta)\right)\\
		\dot{\omega} &= -I^{-1}\omega\times I\omega + I^{-1}\left(m_\rm{a}(\eta,v,\omega,\delta,w)\right)
	\end{align}
	\end{subequations}}
with translational position vector $r(t) \in\mathbb{R}^{3}$, Euler angles $\eta(t) \in\mathbb{R}^{3}$ of the attitude of the bodies and corresponding translational $v(t) \in\mathbb{R}^{3}$ and angular rates $\omega(t) \in\mathbb{R}^{3}$ in the inertia frame of the Earth, where $t\in\mathbb{R}$ denotes time. The input is the left and right tension on the parafoil $\delta(t)\in [0,1]^2$ resulting in its deflection and $w(t)\in\mathbb{R}^3$, which is the wind \chris{velocity that acts as} disturbance. $R:\mathbb{R}^3\rightarrow \mathbb{R}^{3\times 3}$ and $J:\mathbb{R}^3\rightarrow \mathbb{R}^{3\times 3}$ are nonlinear functions of the rotational angles as given in \cite{DeLange2021}. $m$ is the cumulative mass and $I \in \mathbb{R}^{3\times 3}$ is the moment inertia at the center of mass of the spacecraft body, rigidly interconnected with the parafoil. $f_\rm{g}$ is the vectorial gravity force, $f_\rm{a}$ is the aerodynamic force and $m_\rm{a}$ is the aerodynamic moment which are nonlinear functions of the states as detailed in \cite{DeLange2021}. \matthis{The states, forces and moments are all defined in the body frame of the GPRV}. We can write \eqref{eq:under:model}  as
\begin{subequations}\label{NL:model}
\begin{align}
		\dot{x} &= f(x,u,w),\\
		y &= x, \label{NL:model:b}
\end{align}
\end{subequations}
where $x(t)\in\mathbb{R}^{n_\mathrm{x}}$ is the composite state variable in terms of $x=\begin{bmatrix} r^\top & \eta^\top & v^\top & \omega^\top \end{bmatrix}^\top$, with $n_\mathrm{x}=12$, while $u=\delta$ and $w$ is still the wind disturbance. 
 Regarding measurable outputs, we consider the full state vector $y=x$. Furthermore, we can introduce a typical operating region of the system in terms of the compact sets $\mathbb{X}\subset \mathbb{R}^{n_\mathrm{x}}$, $\mathbb{U}\subset \mathbb{R}^{n_\mathrm{u}}$ and $\mathbb{W}\subset \mathbb{R}^{n_\mathrm{w}}$. \chris{In order to give an indication of $\bb{X}$, we have that for a typical trajectory
\begin{align*}
r_{x}(t), r_{y}(t) & \in[-3\cdot 10^3,3\cdot 10^3] \>\>\text{[m]} \\
r_{z}(t) &  \in [6.3\cdot 10^6, 6.4\cdot 10^6] \>\>\text{[m]} \\
\eta(t)&\in[-\pi, \pi]^3 \>\>\text{[rad]} \\
v(t)&\in [-50, 50]^3 \>\>\text{[m/s]}\\
\omega(t)&\in [-0.1, 0.1]^3 \>\>\text{[rad/s],}
\end{align*}
where for $r_{z}$ it must be noted that the radius of the earth is $\sim\matthis{6371}$ km.
Furthermore, we have that $\bb{U}:=[0,1]^2$ and wind disturbances in $\bb{W}$ are considered to be bounded by gusts of 18 [m/s]. Wind data can be obtained via e.g., the National Oceanic and Atmospheric Administration.}
 
\subsection{LPV conversion by factorization}
A common technique to embed a general nonlinear SS model \eqref{NL:model} into an LPV description is to factorize the state transition function $f$ and the output function to obtain
\begin{subequations}\label{eqn:s1:factorization}
\begin{align}
		\dot{x} &= \cal{A}(x,u,w)x+\cal{B}_\mathrm{u}(x,u,w)u+\cal{B}_\mathrm{w}(x,u,w)w,\\
		y &= \cal{C}(x,u,w)x+\cal{D}_\mathrm{u}(x,u,w)u+\cal{D}_\mathrm
		{w}(x,u,w)w,
\end{align}
\end{subequations}
where the matrix functions  $\cal{A},\dots,\cal{D}_\mathrm{w}$ are assumed to be bounded and to have appropriate argument and image dimensions. To obtain \eqref{eqn:s1:factorization} based on \eqref{NL:model}, the aerodynamic forces and moments are factorized w.r.t.~the velocity, angular velocity and the input. The gravity force is factorized w.r.t. the position vector, while due to \eqref{NL:model:b}, $\cal{C}=I$ and  $\cal{D}_\mathrm
		{w}=\cal{D}_\mathrm
		{u}=0$. 

As a second step, a scheduling vector is extracted by constructing the mapping $\theta(t) = \psi(x(t),u(t),w(t)) \in \mathbb{R}^{n_\theta}$ such that the resulting LPV model is
\begin{subequations}\label{eq:lpv_embedding_SR}
 \begin{align} \dot x  & = A(\theta)x + B_\mathrm{u}(\theta)u +  B_\mathrm{w}(\theta)w\\ y &= x.\end{align}
 \end{subequations}
 where $A:\mathbb{R}^{\dnth}\rightarrow \mathbb{R}^{n_\mathrm{x}\times n_\mathrm{x}}$, $B_\mathrm{u}:\mathbb{R}^{\dnth}\rightarrow \mathbb{R}^{n_\mathrm{x}\times n_\mathrm{u}}$ and $B_\mathrm{w}:\mathbb{R}^{\dnth}\rightarrow \mathbb{R}^{n_\mathrm{x}\times n_\mathrm{w}}$ belong to a given function class like affine (that is,
\( A(\theta) = A_0 + \sum_{i=1}^{\dnth} A_i \theta_i \)),  polynomial, etc., and $\mathcal{A}=A\circ \psi$, $\mathcal{B}_\mathrm{u}=B_\mathrm{u}\circ \psi$, $\mathcal{B}_\mathrm{w}=B_\mathrm{w}\circ \psi$, with $\circ$ denoting the function composition operator.

As affine dependence of $A$, $B_\mathrm{u}$ and $B_\mathrm{w}$ is generally preferred for controller design, $\mathcal{A}$, $\mathcal{B}_\mathrm{u}$ and $\mathcal{B}_\mathrm{w}$ is converted to $A\circ \psi$, $B_\mathrm{u}\circ \psi$ and $B_\mathrm{w}\circ \psi$ by extracting every nonlinearity as a new scheduling variable:
  \begin{equation}
 	\theta:=\psi(x, u, w)=\left[\begin{array}{c}
 		f_{1,1}^{A}(x,u,w) \\
 		\vdots \\
 		f_{n_{x}, n_{\mathrm{x}}}^{A}(x,u,w) \\
 		f_{1,1}^{B}(x,u,\matthis{w}) \\
 		\vdots \\
 		f_{n_{\mathrm{x}}, n_{\mathrm{u}}}^{B}(x,u,w) \\
 	\end{array}\right].
 \end{equation}
 As a last step, the scheduling region $\Theta$ as $\psi(\mathbb{X},\mathbb{U},\mathbb{W})\subseteq \Theta$ is computed, where $\Theta$ is taken as a the smallest hypercube based on the extreme values of each component of $\psi$ over $(\mathbb{X},\mathbb{U},\mathbb{W})$.
 
 While this constitutes to a simple LPV model conversion process where the obtained model is an exact representation of the original nonlinear system,  the conservativeness and complexity of the representation are maximized, achieving a scheduling dimension $\dnth=71$. As a next step, we will reduce $\dnth$ and optimize the conservativeness of the LPV representation of the \matthis{GPRV} dynamics. Furthermore, we will show that the wind $w$ can be excluded from $\psi$ without significant deterioration of the model accuracy.
\section{Scheduling reduction methods}\label{s:prelim}
For reducing the conservativeness and complexity of the converted LPV model obtained in Section \ref{sec:SR:mod}, we will briefly introduce the PCA method developed in \cite{sadeghzadeh2020affine} and the DNN approach from \cite{Koelewijn2020} in this section. These methods will be applied to the \matthis{GPRV} LPV model in Section \ref{sec:sr_of_sr}.

\subsection{PCA-based scheduling dimension reduction}\label{ss:PCA}
The PCA-based scheduling dimension reduction method is based on \cite{sadeghzadeh2020affine}, which is an improved version of \citep{Kwiatkowski2008}. The idea of the PCA method is to extract dominant components,  i.e., the \emph{principle components}, of the model variations that contribute most to the system behavior along typical operational trajectories of the system. These principle components are extracted by means of a \emph{Singular Value Decomposition} (SVD), which allows to determine an effective number of components on which the new model can be scheduled. 

In order to apply the PCA approach, it is assumed that variations of the system induced by typical operational trajectories are given in terms of a data-set $\Pi_N$ on which the PCA is performed. Given a set of state and input points $\mathcal{D}_N=\{{x}(k),{u}(k),{w}(k)\}_{k=1}^{N}\in\mathbb{R}^{n_\mathrm{x}\times n_\mathrm{u}}$ that are sampled from multiple typical flight trajectories of the \matthis{GPRV}, the corresponding ${\theta}(k)\in\Theta_N=\psi(\mathcal{D}_N)$ scheduling variation, and the variation of the matrices $A(\theta(k)), \dots, D_\matthis{\rm{w}}(\theta(k))$ reshaped as  
\begin{align}\label{eq:vec}
	\Gamma(\theta(k)) &= \rm{vec}\left(\begin{bmatrix}
		A(\theta)&B_\mathrm{u}(\theta) & B_\mathrm{w}(\theta) \\
		C(\theta)&D_\mathrm{u} (\theta) & D_\mathrm{w}(\theta)
	\end{bmatrix}\!\!(k)\right), 
	\end{align}
where $\rm{vec}$ denotes column vectorization of a matrix. The considered trajectories $\mathcal{D}_N$ should be chosen that they represent the solution space that the nonlinear model encounters during typical operation of the system. Then, the  variational data-set $\Pi_N$ is obtained as
\begin{equation}
	\Pi_N = \begin{bmatrix} \Gamma(\theta(1)) & \Gamma(\theta(2)) & \cdots  & \Gamma(\theta(N)) \end{bmatrix} \in \mathbb{R}^{n_{\Pi}\times N}.	
\end{equation} 
where $n_{\Pi}=(n_\mr{x}+n_\mr{u})(n_\mr{x}+n_\mr{y})$.

To improve numerical conditioning, the data is often centered and normalized 
\begin{equation}
\bar{\Pi}_N = \normal(\Pi_N):= S_\mathrm{scale} \cdot (\Pi_N-\Pi_\mathrm{c} \otimes 1_{ n_{\Pi} \times N} ).
\end{equation}
where $\Pi_\mathrm{c}\in\mathbb{R}^{n_{\Pi}}$ is the column average, i.e, mean, of $\Pi_N$, $S_\mathrm{scale}\in\mathbb{R}^{n_{\Pi} \times n_{\Pi}}$ is a diagonal scaling matrix \matthis{and $\otimes$ denotes the Kronecker product}.

\begin{rem}\label{rem:norm}
    Normalization can be accomplished in terms of the standard deviation and min-max normalization. For the standard deviation based normalization 
       \begin{equation}\label{eq:nrmdef}
        S_\mathrm{scale}:= \mathrm{diag}^{-1}(\mathrm{std}(\Pi_{N,1}), \ldots \mathrm{std}(\Pi_{N,n_\Pi}))
    \end{equation}
    where $\rm{std}$ is the square root of the sample variance and $\Pi_{N,i}$ corresponds to the $i^\rm{th}$ row of $\Pi_N$. For min-max normalization,  $S_\mathrm{scale}$ is defined in terms of 
           \begin{equation}
        S_\mathrm{scale}:= \mathrm{diag}^{-1}(d(\Pi_{N,1}), \ldots, d(\Pi_{N,n_\Pi}))
    \end{equation}
    where $d(\Pi_{N,i})=\max(\Pi_{N,i})-\min(\Pi_{N,i})$.  This means that the data is scaled and centered between -1 and 1 with 0 mean.
     \hfill $\square$
\end{rem}

We can obtain $\Pi_N$ from $\bar\Pi_N$ with the inverse scaling and centering operation denoted $\normal^{-1}$.
The principle components are extracted from $\bar{\Pi}_N$ with an SVD, i.e., 
\begin{equation}
	\bar{\Pi}_N = U\Sigma V^{\top} = \begin{bmatrix}
		U_{\rm{s}}&U_{\rm{r}}
	\end{bmatrix}
	\begin{bmatrix}
		\Sigma_\rm{s}&0\\
		0&\Sigma_\rm{r}
	\end{bmatrix}
	\begin{bmatrix}
		V_\rm{s}^{\top}\\
		V_\rm{r}^{\top}
	\end{bmatrix},
\end{equation}
with $U$ and $V$ orthonormal matrices containing the singular vectors. $\Sigma$ is a diagonal matrix with the singular values of $\Pi_N$ on the diagonal in descending order. 
Projecting the data-set to a lower dimension, 
while retaining the most {significantly contributing} varying components, is done by taking the first $n_{\rm{s}}$ principle components, i.e., singular values, of $\Pi_N$. This results in
\begin{equation}
\hat{\bar{\Pi}}_N = U_{\rm{s}}\Sigma_\rm{s}V_\rm{s}^{\top}= U_{\rm{s}}U_{\rm{s}}^{\top}\bar{\Pi}_N.
\end{equation} 
The core idea is to use $U_{\rm{s}}^{\top}\bar{\Pi}_N$ as the new scheduling map, whose dimension is equal to the selected principal components, i.e., $n_\mr{s}=n_{\hat{\theta}}$.  Here $U_{\rm{s}}$ describes the linear combination of these variables which describe the variation, i.e., how these new scheduling variables will compose a new affine dependency of the LPV model. Note that such a decomposition is based on normalized and centered variations, hence the approximation of the original matrix variations is found with the inverse of the normalization $\cal{N}^{-1}(\hat{\bar{\Pi}}_N)=S_\mathrm{scale}^{-1} \hat{\bar{\Pi}}_N + \Pi_\mathrm{c} \otimes 1_{n_{\Pi} \times N} $, which is still an affine operator, preserving the affine dependency structure of the LPV model. Based on these, the reduced scheduling variable $\hat{\theta}$ is given as
\begin{equation}
	\hat{\theta}(t) \!=\! U_{\rm{s}}^{\top}\!\normal(\Gamma(\theta(t)))\!=\! \underbrace{U_{\rm{s}}^{\top}\!\normal(\Gamma(\psi(x(t),u(t),w(t))))}_{\hat{\psi}(x(t),u(t),w(t))}.
\end{equation}
The new linear affine mapping from $\hat{\theta}$ to approximate the original $\mathcal{A}, \dots, \mathcal{D}_\mathrm{w}$ is reconstructed from the approximated model variations $\hat{\Gamma}(\hat{\theta}) = \normal^{-1}(U_{\rm{s}}\hat{\theta})$ and gives
\begin{multline}\label{eqn:PCA:Mhat}
\begin{bmatrix}
	{A}(\theta)&{B}_\matthis{\rm{u}}(\theta)&{B}_\matthis{\rm{w}}(\theta)\\
	{C}(\theta)&{D}_\matthis{\rm{u}}(\theta)&{D}_\matthis{\rm{w}}(\theta)
\end{bmatrix} \approx  \begin{bmatrix}
		\hat{A}(\hat\theta)&\hat{B}_\matthis{\rm{u}}(\hat\theta)&\hat{D}_\matthis{\rm{u}}(\hat\theta)\\
		\hat{C}(\hat\theta)&\hat{D}_\matthis{\rm{u}}(\hat\theta)&\hat{D}_\matthis{\rm{w}}(\hat\theta)
	\end{bmatrix} = \\
	= \hat{M}(\hat\theta) = \hat{M}_0 + \sum_{i = 1}^{n_{\hat{\theta}}}\hat{\theta}_i\hat{M}_i,
\end{multline}
with $n_{\hat{\theta}} < n_{\theta}$, and where $\hat{M}_0 = \begin{bmatrix}{A}_0 & {B}_\matthis{\rm{u,0}} & {B}_\matthis{\rm{w,0}}\\ {C}_0 & {D}_\matthis{\rm{u,0}} & {D}_\matthis{\rm{w,0}}\end{bmatrix} + \mr{vec}^{-1}(\Pi_\mathrm{c})$ and $\hat{M}_i =  \mr{vec}^{-1}(S_\mathrm{scale}^{-1} U_{\mr{s},i})$ with $\mathrm{vec}^{-1}$ the inverse operation of \eqref{eq:vec} and $U_{\mr{s},i}$ representing the $i$\textsuperscript{th} column of $U_{\mr{s}}$ with $1\leq i\leq n_\mr{s}$. 

\chris{The final step in the PCA scheduling reduction method is to determine the reduced scheduling region $\hat{\Theta}$ in which $\hat{\theta}$ is varying. The region $\hat{\Theta}$ can be defined as a hypercube denoted as 
\[\hat{\theta}_i^{\rm{min}}\le \hat\theta_i\le\hat{\theta}_i^{\rm{max}}, \]
where $\hat{\theta}_i^{\rm{min}}$ and $\hat{\theta}_i^{\rm{max}}$ are obtained as the minimum and maximum values of  $\hat{\theta}_i(t)$, respectively, over all admissible values of $\theta\in\psi(\bb{X}, \bb{U})$. This however is often not the hypercube with the smallest volume, which introduces conservatism. In \cite{sadeghzadeh2020affine}, the problem of finding this minimum-volume region for $\hat{\Theta}$ is discussed for the cases where $n_{\hat{\theta}}\le 3$ and where $n_{\hat{\theta}}> 3$. The former case makes use of the Kabsch algorithm \citep{kabsch1976solution}, while the latter aims to find the minimum-volume hyper-ellipse that encloses the trajectories of $\hat{\theta}$ generated with $\cal{D}_N$. The principle axes of the hyper-ellipse are then used to construct the hypercube describing $\hat{\Theta}$. See \cite[Sec. 4]{sadeghzadeh2020affine} for details on both methods.}

\subsection{DNN-based scheduling dimension reduction}
\begin{figure}[!ht]
\centering
\includegraphics[width=0.7\linewidth]{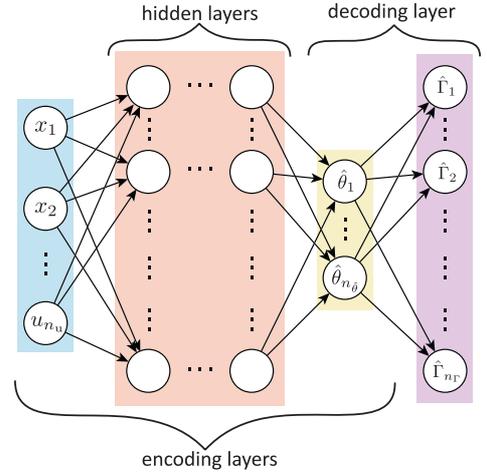}
\caption{DNN architecture for learning-based scheduling reduction (adopted from \cite{Koelewijn2020}). 
}
\label{fig:Pre:DNN}
\end{figure}

The PCA method uses a linear mapping from a fixed set of model variations governed by the scheduling $\theta$ to the reduced scheduling vector $\hat\theta$. With the DNN method, proposed in \cite{Koelewijn2020}, {the scheduling map $\psi$ {is} learned, i.e. optimized, simultaneously along the reduction step. This implies that the model} accuracy {with} reduced scheduling can be possibly improved by exploiting a direct but compact nonlinear mapping from $(x,u)$ to the reduced scheduling vector. 
In Figure~\ref{fig:Pre:DNN} a schematic overview of the DNN architecture is given which is used in this reduction method. The DNN encodes the states and the inputs into the reduced scheduling vector and the linear layer decodes the reduced scheduling vector to the approximated model variations.
The DNN consists of an input layer, $n_\rm{hl}$ hidden layers and an output layer. {The input and hidden layers {are} described as 
\begin{equation}
	l^{[\tau]} = g^{[\tau]}\left(W^{[\tau]}l^{[\tau-1]}+b^{[\tau]}\right), \quad \tau = 0, \dots, n_\rm{hl}, 
\end{equation} 
where $g^{[\tau]}(\cdot)$ is the activation function \chris{(such as a hyperbolic tangent, rectified linear unit (ReLU), sigmoid, etc. function)}
, $l^{[\tau]}$ is the output of the $\tau$\tss{th} layer and $W^{[\tau]}$ and $b^{[\tau]}$ are the weighting matrix and bias vector of the $\tau$\tss{th} layer. The input to the DNN, i.e., $(x,u)$, is thus $l^{[-1]}:=\rm{vec}(x,u)$. 
%
The reduced scheduling vector is the output of the $n_\rm{hl}$\tss{th} hidden layer, i.e., $\hat\theta := l^{[n_\rm{hl}]}$. The associated vectorized system matrices $\hat{\Gamma}$ follow from the last layer, which is affine,}
\begin{equation}
	\hat{\Gamma} = W^{[\Gamma]}\hat{\theta}+b^{[\Gamma]}.
\end{equation}
The relation from the reduced scheduling vector to the approximated matrices $\hat{A}$, $\hat{B}$, $\hat{C}$ and $\hat{D}$ as given in \eqref{eqn:PCA:Mhat} directly follows from the last layer, taking the inverse of the data normalization into account. The weightings and biases of the DNN are optimized by minimizing 
\begin{equation}\label{eq:lossfunc}
	\min_{W^{[k]},b^{[k]}} \frac{1}{N}\sum_{j=1}^{N}\left|\left|\hat{\Gamma}(\hat{\theta}(j))-\Gamma(\theta(j))\right|\right|^2_2. 
\end{equation} 
The optimization problem is solved with a back-propagation algorithm combined with a stochastic gradient descent, which are implemented in popular solvers  such as Adam, or AdaBound \citep{kingma2014adam, luo2019adaptive}. Multiple techniques exist to prevent overfitting, like weight regularization and {early stopping} \citep{ref:bookonDeepLearning}.

\chris{Based on the obtained $\hat{\psi}, \hat{\theta}$, the scheduling region $\hat{\Theta}$ can be determined using the same methods as discussed in Section~\ref{ss:PCA}}


\section{Reduced LPV model of the GPRV}\label{sec:sr_of_sr}
Armed with the PCA and DNN methods introduced in Section \ref{s:prelim}, we optimize the scheduling complexity in the LPV modeling of the  flight dynamics together with the conservativeness of the embedding. 
\subsection{Data-generation}
We perform the optimization of the complexity and conservativeness based on trajectory data from typical operation of the \matthis{GPRV}. SENER Aerospace presented in \cite{Aerospace2019a} a baseline solution of the GNC problem on a simplified model of the \matthis{GPRV}. From the associated simulator, we obtained typical initial conditions for the states and input trajectories of $\delta$ to simulate our high-fidelity model (as presented in Section~\ref{ss:highfidmodel}) in open-loop. These trajectories navigate the \matthis{GPRV} from $\sim$5.5 km above the Earth's surface to a predefined landing location. \todo{As the wind cannot be measured during operation, we are not able to schedule the model based on the wind. For this reason, we exclude the wind in the reduction techniques, i.e., $w=0$, and we assume that the GNC will be able to reject the disturbance in closed-loop control}. The simulation is computed with an ODE4 solver, with a sampling frequency of 400 Hz. The resulting data-set has a total of over $10^7$ data-points.

\subsection{Optimizing the direct LPV model of the GPRV}\label{ss:optres}
We will now further optimize the direct LPV model \eqref{eq:lpv_embedding_SR} in both scheduling complexity and conservativeness. The former will be accomplished with the PCA and DNN-based scheduling dimension reduction methods. The latter comes along with the construction of the scheduling region, which we will only discuss briefly.
\subsubsection{\ref{ss:optres}.1~PCA-based scheduling construction:}
From \eqref{eq:lpv_embedding_SR} and the design choice that $w=0$, we can define $\Gamma(\theta(k))$ as
\begin{equation}
\Gamma(\theta(k)) = \rm{vec}\left(\begin{bmatrix} A(\theta)&B_\mathrm{u}(\theta) \end{bmatrix}\!(k)\right),
\end{equation}
from which we generate the data-set $\Pi_N$. As we will need to compute the SVD of $\Pi_N$, we use $N= 5\cdot 10^{5}$ points to generate it, which are randomly sampled from the $10^7$-point data-set, to retain computational tractability. We normalize $\Pi_N$ with both $\mr{std}$ and $\min$-$\max$ normalization (see Remark~\ref{rem:norm}).
\begin{figure}[!ht]
\centering
\includegraphics[width=.7\linewidth]{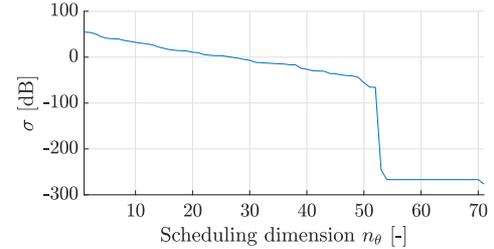}
\caption{The singular values of $\bar{\Pi}_N$ associated with the PCA, where $\bar{\Pi}_N$ has been normalized with min-max normalization. }
\label{fig:PCA:sv}
\end{figure}
The principle components of $\bar{\Pi}_N$, i.e., the singular values, with min-max normalization are plotted in Figure \ref{fig:PCA:sv}, which shows an exponential decrease of the singular values. For the 52\textsuperscript{nd} singular value till the last, the singular values drop below the numerical precision bound and can be considered zero. Hence, the LPV embedding \eqref{eq:lpv_embedding_SR} requires 52 principal components to describe the variations of $\mathcal{A}, \mathcal{B}_\mr{u}$ along typical operation of the \matthis{GPRV}. We now construct a compact scheduling map to economically represent the \matthis{GPRV} in an LPV form, which results in making a trade-off between complexity of the LPV model and model accuracy.
In order to visualize this trade-off, we compute the approximated data matrix $\hat{\Pi}$ for a scheduling dimension of $n_{\hat{\theta}}= 1, \dots, 10$. This is often the range that is numerically manageable in controller synthesis problems for systems with $n_\mr{x}>10$.
\subsubsection{\ref{ss:optres}.2~DNN-based scheduling construction:}
We will now optimize the direct LPV model in terms scheduling complexity for a scheduling dimension of $n_{\hat{\theta}}= 1, \dots, 10$ using the DNN-based tools. The DNN is implemented with 4 hidden layers, each consisting of 128 neurons with $\tanh$ activation. Moreover, we applied a linear bypass between the input and output. The DNN input is $\begin{bmatrix} r^\top & \tilde{\eta}^\top & V^\top & \omega^\top & \delta^\top \end{bmatrix}^\top$, where $\tilde{\eta}^\top:= \begin{bmatrix} \sin(\eta)^\top & \cos(\eta)^\top \end{bmatrix}^\top$, resulting in 17 inputs. This decomposition of the angular states often helps in the training of the network. The decoding layer, i.e., the output layer of the DNN (as depicted in Figure~\ref{fig:Pre:DNN}), has $n_{\hat{\theta}}= 1, \dots, 10$ inputs\footnote{Note that the network has to be retrained for every $n_{\hat{\theta}}$.} and the $n_\Pi=71$ model variations as output. Note that this is a linear layer that results in the linear affine mapping from the reduced scheduling vector to the approximated matrix functions $(\hat{A},\hat{B}_\matthis{\rm{u}})$. The $\ell_2$-weight regularization is set to $10^{-4}$. 
The Adam optimizer \cite[]{kingma2014adam} is used during training with a learning rate of $10^{-5}$. The batch-size is 128 and the network is trained to minimize \eqref{eq:lossfunc} for 200 epochs. Both the input and output data are normalized before training, with the aforementioned normalization methods. We want to stress here that the DNN-based  scheduling reduction approach simultaneously finds a nonlinear map between the input $(x,u)$ and a scheduling vector of size $n_{\hat{\theta}}$, and an affine map between the scheduling vector and the model variations.

\subsection{Comparison of the results}\label{ss:optres2}
\begin{figure}[t]
\centering
\begin{subfigure}[b]{\linewidth}
\includegraphics[width=\linewidth]{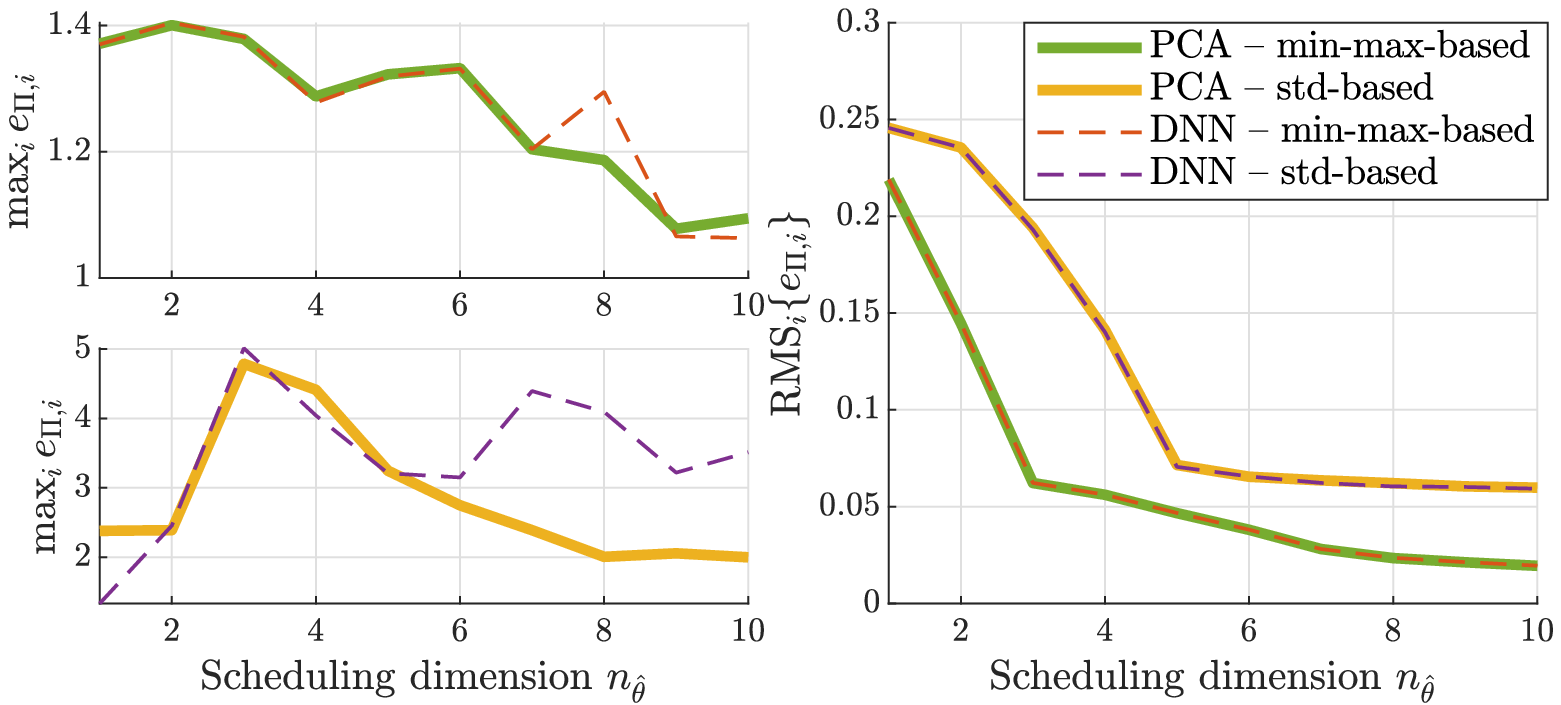}
\caption{Comparison using the error measure in \eqref{eq:meas1}.}
\label{fig:PCA:compare}
\end{subfigure}
\begin{subfigure}[b]{\linewidth}
\centering
\includegraphics[width=\linewidth]{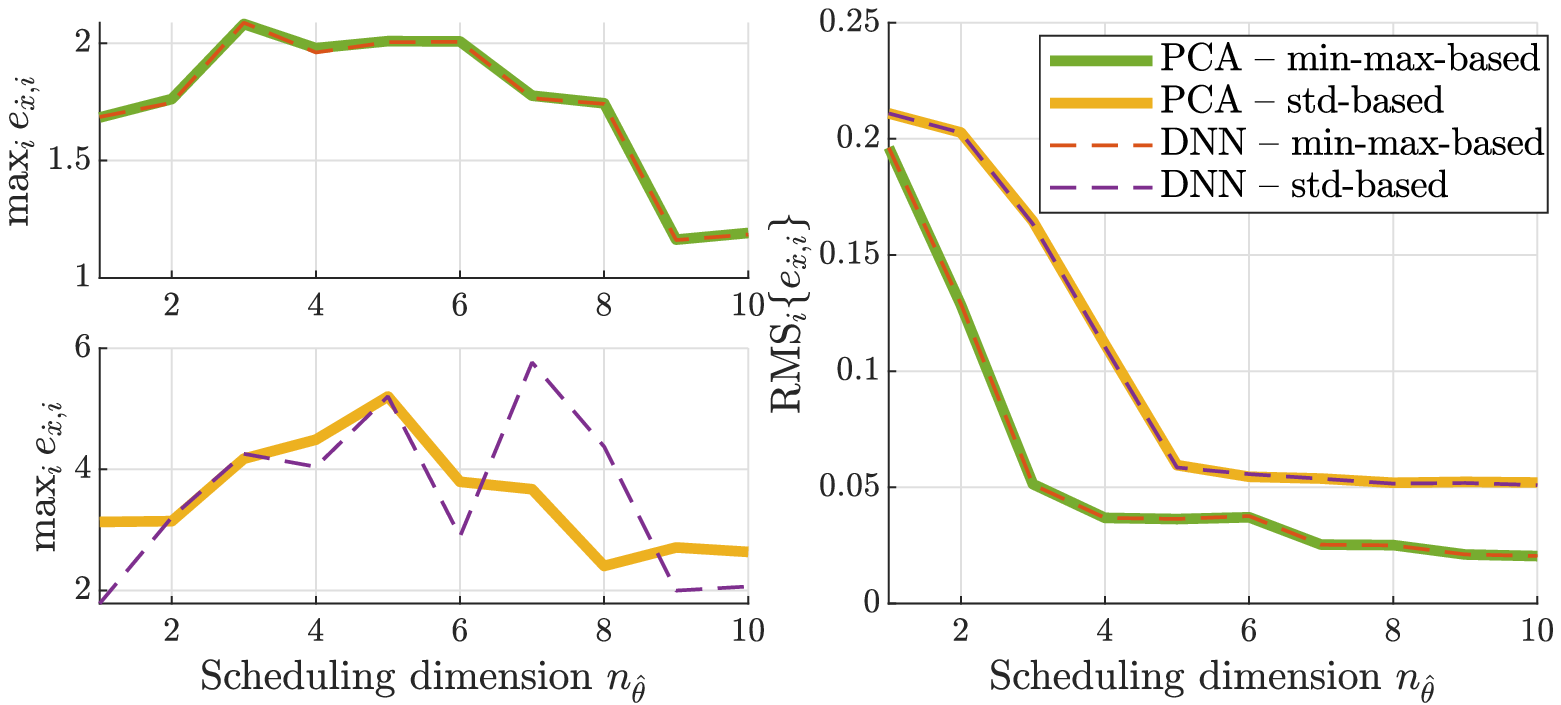}
\caption{Comparison using the error measure in \eqref{eq:meas2}.} 
\label{fig:MP:com}
\end{subfigure}
\caption{Comparison of the results on scheduling complexity optimization. The results are plotted for both the PCA (solid lines) and the DNN (dashed lines) approach, with the cases of min-max-based and std-based normalization of the data.}\label{fig:results1}
\end{figure}

\begin{figure}[!ht]
\centering
\includegraphics[width=\linewidth]{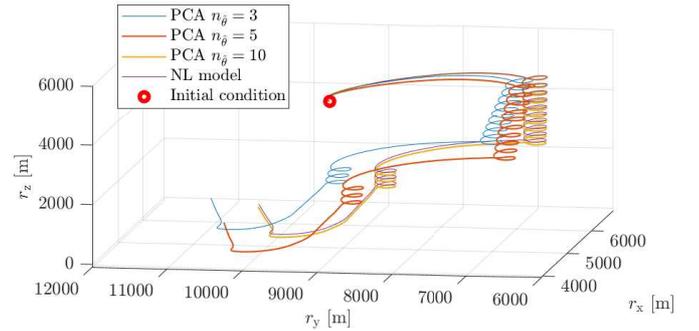}
\caption{Flight trajectory of the full nonlinear model of the \matthis{GPRV}, compared to the reduced LPV models with $n_{\hat\theta}=\{3,5,10\}$.}
\label{fig:trajectories}
\end{figure}

We compare the outcomes of the scheduling reduction methods using two types of error measures on a validation data-set, which is uncorrelated from the training data-set. The first error measure is the normalized approximation error of the elements of $\Gamma$, i.e., along the rows of $\Pi$. Let
\begin{equation}\label{eq:meas1}
e_{\Pi, i} := \frac{\|\Pi_{N,i}-\hat{\Pi}_{N,i}\|_2}{\left\|\Pi_{N,i}\right\|_\infty}, \quad i = 1, \dots, n_\Pi,
\end{equation}
with $\|\cdot\|_2$ and $\|\cdot\|_\infty$ the Euclidean $\ell_2$ and $\ell_\infty$ vector-norms, respectively.
While this error measure gives a good indication how $\begin{bmatrix} \hat{A}(\hat{\theta}) & \hat{B}_{\matthis{\rm{u}}}(\hat{\theta}) \end{bmatrix}$ characterizes $\begin{bmatrix} \cal{A}(x, u) & \cal{B}_{\matthis{\rm{u}}}(x, u)\end{bmatrix}$, we are mainly interested in how well the obtained LPV model represents the \emph{true} solution space of the \matthis{GPRV}. A much closer characterization for this is comparing the state-derivatives, which compares $\cal{A}(x,u)x+\cal{B}_\mr{u}(x,u)u$ with $\hat{A}(\hat{\theta})x+\hat{B}_{\matthis{\rm{u}}}(\hat{\theta})u$. Let
\begin{align*}
f^N &\! = \begin{bsmallmatrix} {A}(\theta(1))x(1)+{B}_\mr{u}(\theta(1))u(1) & \cdots & {A}(\theta(N))x(N)+{B}_\mr{u}(\theta(N))u(N) \end{bsmallmatrix}, \\
\hat{f}^N & \! = \begin{bsmallmatrix} \hat{A}(\hat\theta(1))x(1)+\hat{B}_\mr{u}(\hat\theta(1))u(1) & \cdots & \hat{A}(\hat\theta(N))x(N)+\hat{B}_\mr{u}(\hat\theta(N))u(N) \end{bsmallmatrix},
\end{align*}
where $f^N,\hat{f}^N\in\bb{R}^{n_\mr{x}\times N}$. With this, the second error measure is defined as
\begin{equation}\label{eq:meas2}
e_{\dot{x}, i} := \frac{\|f^N_{i}-\hat{f}^N_i\|_2}{\left\|f^N_{i}\right\|_\infty}, \quad i = 1, \dots, n_\mr{x}.
\end{equation}
Figure~\ref{fig:results1} shows the plots for $\max_i e_{\Pi, i}$, $\rm{RMS}_i\{e_{\Pi, i}\}$, $\max_i e_{\dot{x}, i}$ and $\rm{RMS}_i\{e_{\dot{x}, i}\}$, where $\rm{RMS}_i\{a_i\}:= \sqrt{\sum_{i=1}^{n_\mr{a}}a_i^2}$, for both the PCA and DNN-based scheduling reduction methods. 
The plots show that there is no significant difference between the PCA and the DNN for both the error measures. The main property that causes a difference in the result is the type of normalization used on the data, where the min-max normalization clearly shows better results. It must be noted that a clear advantage of the DNN method is that we have direct control over which system variables are participating in the new scheduling map, while there is no control over this for the PCA-based approach. Moreover, when considering min-max-based normalization, a reduced scheduling dimension of $n_{\hat\theta}=3$ would be the optimal trade-off between complexity and accuracy, as the RMS error does not get significantly smaller for $n_{\hat\theta}>3$. When we compare the behavior during a nominal flight trajectory, we obtain the trajectories in Figure~\ref{fig:trajectories}.
We want to highlight here that the trajectories are simulated in open-loop, hence the deviation over time will likely vanish in the case of closed-loop operation.

\subsubsection{\ref{ss:optres2}.1~Optimizing the conservativeness:} 
\begin{figure}[t]
\centering
\begin{subfigure}[b]{.45\linewidth}
\centering
\includegraphics[width=\linewidth]{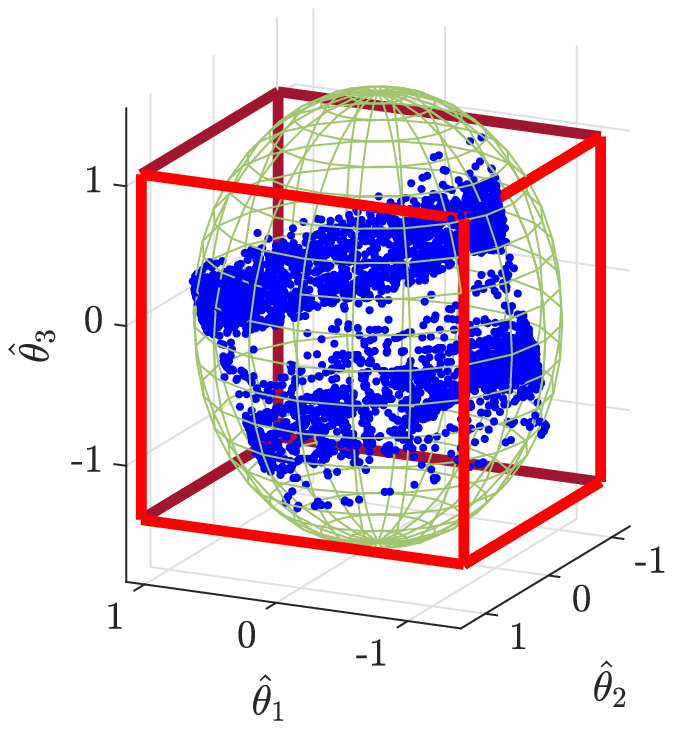}
\caption{$\hat\Theta$ from PCA method.}
\label{fig:schedulingrange:PCA}
\end{subfigure}~~
\begin{subfigure}[b]{.45\linewidth}
\centering
\includegraphics[width=\linewidth]{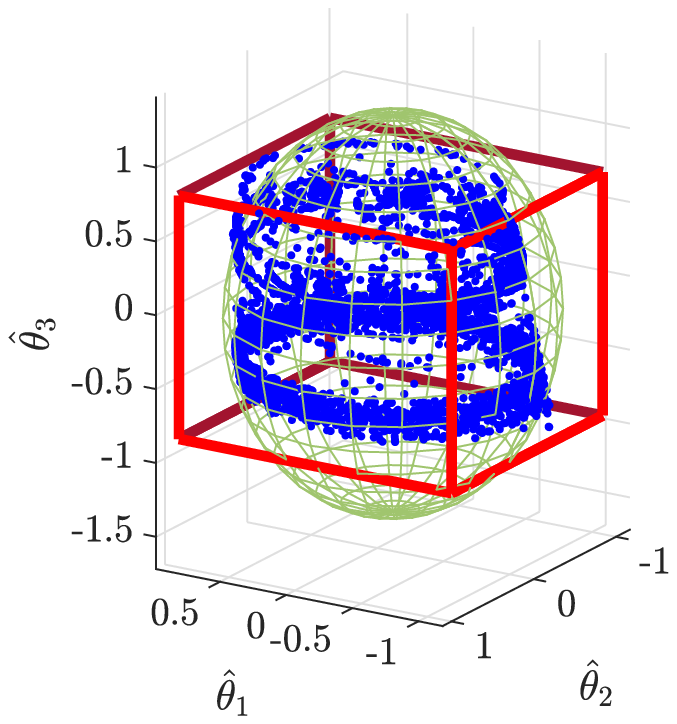}
\caption{$\hat\Theta$ from DNN method.}
\label{fig:schedulingrange:DNN}
\end{subfigure}
\caption{Construction of $\hat\Theta$, with the data-points $\hat\theta(k)$ in blue, the minimum-volume sphere in green and the cube defining $\hat\Theta$ in red.}
\label{fig:schedulingrange}
\end{figure}

We will briefly discuss the construction of $\hat{\Theta}$ for the case where $n_{\hat\theta}=3$. The scheduling range for $\hat\theta$ is constructed by finding the minimum-volume sphere that encloses $U_{n_{\hat\theta}}^{\top}\bar\Pi_N$, which contains all the points of $\hat\theta(k)$ applied on the validation data-set. Using the methodology described in \cite[Sec. 4.2]{sadeghzadeh2020affine}, we obtain the cube that defines $\hat\Theta$, which is depicted in Figure~\ref{fig:schedulingrange}.
To give an indication of the conservatism, we calculate the ratio between `un-used' and `used' space of $\hat\Theta$, i.e., $\tfrac{\text{Volume(cube)}-\text{Volume(polytope)}}{\text{Volume(polytope)}}$, with Volume(cube) the volume of the cube, and Volume(polytope) the volume of the convex polytope around $\hat\theta(k)$. The resulting ratios are 1.05 for the PCA based optimization and 0.58 for the DNN-based optimization for $n_{\hat{\theta}}=3$, which implies that the PCA method is slightly more conservative.
\subsubsection{\ref{ss:optres2}.2~Discussion}
From the results we can conclude that we can obtain an affine LPV embedding of the complex nonlinear \matthis{GPRV} system with a scheduling dimension of 3. Such a level of scheduling complexity can usually easily be handled in controller synthesis problems. 
In case the system is exactly known, the PCA-based complexity optimization method would be the least conservative approach. If this is not the case, the DNN would be more favorable, as this method directly learns the scheduling map $\hat\psi(x(t),u(t)):=\hat{\theta}(t)$, which can be easily implemented in a GNC design. 

We want to remark a final note on the interpretability of the reduced scheduling maps. In aerospace applications it is often desired to have a physical understanding of the scheduling variables. In our case, it is very difficult and possibly not even possible to preserve the physical interpretation of $\hat\theta$, which is the trade for having an accurate model that can be used for the synthesis of an high-performance LPV controller.

\section{Conclusions} \label{sec:conclusion}


This paper presents the LPV conversion process of \matthis{a} nonlinear \matthis{Generic Parafoil Return Vehicle}, where we optimize the LPV model over its complexity and conservativeness, such that we obtain a model that is suitable for control synthesis. For this optimization, we applied the PCA and DNN-based scheduling dimension reduction techniques. Our results show that the highly complex nonlinear model of the \matthis{GPRV} benchmark system can be embedded in an LPV representation with a scheduling dimension of 3. The DNN-based method simultaneously learns a scheduling map, which is favorable from an implementation point of view. 
For future work, we aim to design an high-performance LPV controller for the reduced system and test its performance in closed-loop with the high-fidelity nonlinear model.


\begin{ack}
We gracefully thank the SENER group and P.J.W. Koelewijn for useful discussions and providing simulation tools of the GPRV.
\end{ack}

\bibliography{Master_thesis_Prep_report}

\begin{thebibliography}{25}
\providecommand{\natexlab}[1]{#1}
\providecommand{\url}[1]{\texttt{#1}}
\providecommand{\urlprefix}{URL }
\expandafter\ifx\csname urlstyle\endcsname\relax
  \providecommand{\doi}[1]{doi:\discretionary{}{}{}#1}\else
  \providecommand{\doi}{doi:\discretionary{}{}{}\begingroup
  \urlstyle{rm}\Url}\fi

\bibitem[{Beck(2006)}]{beck2006modelbasedred_lpv_coprime}
Beck, C. (2006).
\newblock Coprime factors reduction methods for linear parameter varying and
  uncertain systems.
\newblock \emph{Systems \& Control Letters}, 55(3), 199--213.

\bibitem[{Cacciatore et~al.(2019)Cacciatore, Ramos, Castellani, Figueroa,
  Veenman, Ram{\'{i}}rez, Recupero, Kerr, and B{\'{e}}jar}]{Aerospace2019a}
Cacciatore, F., Ramos, H.R., Castellani, T.L., Figueroa, A., Veenman, A.,
  Ram{\'{i}}rez, S., Recupero, C., Kerr, M., and B{\'{e}}jar, J. (2019).
\newblock {The Design of the GNC of the Re-entry Module of Space Rider}.
\newblock In \emph{Proc. of the 8\tss{th} European Conference for Aeronautics
  and Space Sciences}.

\bibitem[{Carter and Shamma(1996)}]{Carter1996}
Carter, L.H. and Shamma, J.S. (1996).
\newblock {Gain-scheduled bank-to-turn autopilot design using linear parameter
  varying transformations}.
\newblock \emph{Journal of Guidance, Control, and Dynamics}, 19(5), 1056--1063.

\bibitem[{Corti et~al.(2012)Corti, Dardanelli, and Lovera}]{corti2012lpv}
Corti, A., Dardanelli, A., and Lovera, M. (2012).
\newblock {LPV methods for spacecraft control: An overview and two case
  studies}.
\newblock In \emph{Proc. of the American Control Conference}, 1555--1560.

\bibitem[{de~Lange(2021)}]{DeLange2021}
de~Lange, M. (2021).
\newblock {Modeling of the Space Rider Flight Dynamics During the Terminal
  Descent Phase}.
\newblock Technical report, University of Technology Eindhoven.

\bibitem[{Figueroa-Gonz{\'{a}}lez et~al.(2021)Figueroa-Gonz{\'{a}}lez,
  Cacciatore, and Haya-Ramos}]{Figueroa-Gonzalez2021a}
Figueroa-Gonz{\'{a}}lez, A., Cacciatore, F., and Haya-Ramos, R. (2021).
\newblock {Landing Guidance Strategy of Space Rider}.
\newblock \emph{Journal of Spacecraft and Rockets}, 58(4), 1220--1231.

\bibitem[{Goodfellow et~al.(2016)Goodfellow, Bengio, and
  Courville}]{ref:bookonDeepLearning}
Goodfellow, I., Bengio, Y., and Courville, A. (2016).
\newblock \emph{Deep Learning}.
\newblock MIT Press.

\bibitem[{Hecker and Varga(2005)}]{hecker2005symbolic}
Hecker, S. and Varga, A. (2005).
\newblock Symbolic techniques for low order lft-modelling.
\newblock In \emph{Proc. of the 16\tss{th} IFAC World Congress}, volume~38,
  523--528.

\bibitem[{Hoffmann and Werner(2015)}]{Hoffmann2015LPV_Gyro}
Hoffmann, C. and Werner, H. (2015).
\newblock {LFT-LPV modeling and control of a Control Moment Gyroscope}.
\newblock In \emph{Proc. of the 54\tss{th} IEEE Conference on Decision and
  Control}, 5328--5333.

\bibitem[{Hoffmann and Werner(2014)}]{hoffmann2014survey}
Hoffmann, C. and Werner, H. (2014).
\newblock A survey of linear parameter-varying control applications validated
  by experiments or high-fidelity simulations.
\newblock \emph{IEEE Transactions on Control Systems Technology}, 23(2),
  416--433.

\bibitem[{Kabsch(1976)}]{kabsch1976solution}
Kabsch, W. (1976).
\newblock A solution for the best rotation to relate two sets of vectors.
\newblock \emph{Acta Crystallographica: Section A}, 32(5), 922--923.

\bibitem[{Kingma and Ba(2014)}]{kingma2014adam}
Kingma, D.P. and Ba, J. (2014).
\newblock Adam: A method for stochastic optimization.
\newblock \emph{arXiv preprint arXiv:1412.6980}.

\bibitem[{Koelewijn and T{\'o}th(2020)}]{Koelewijn2020}
Koelewijn, P.J.W. and T{\'o}th, R. (2020).
\newblock {Scheduling Dimension Reduction of LPV Models-A Deep Neural Network
  Approach}.
\newblock In \emph{Proc. of the American Control Conference}, 1111--1117.

\bibitem[{Kwiatkowski et~al.(2006)Kwiatkowski, Boll, and
  Werner}]{Kwiatkowski2006Automated_generation}
Kwiatkowski, A., Boll, M.T., and Werner, H. (2006).
\newblock {Automated Generation and Assessment of Affine LPV Models}.
\newblock In \emph{Proc. of the 45\tss{th} IEEE Conference on Decision and
  Control}, 6690--6695.

\bibitem[{Kwiatkowski and Werner(2008)}]{Kwiatkowski2008}
Kwiatkowski, A. and Werner, H. (2008).
\newblock {PCA-based parameter set mappings for LPV models with fewer
  parameters and less overbounding}.
\newblock \emph{IEEE Transactions on Control Systems Technology}, 16(4),
  781--788.

\bibitem[{Luo et~al.(2019)Luo, Xiong, Liu, and Sun}]{luo2019adaptive}
Luo, L., Xiong, Y., Liu, Y., and Sun, X. (2019).
\newblock Adaptive gradient methods with dynamic bound of learning rate.
\newblock \emph{arXiv preprint arXiv:1902.09843}.

\bibitem[{Marcos and Balas(2004)}]{Marcos2004}
Marcos, A. and Balas, G.J. (2004).
\newblock {Development of Linear Parameter-Varying Models for Aircraft}.
\newblock \emph{Journal of Guidance, Control, and Dynamics}, 27(2), 218--228.

\bibitem[{Rizvi et~al.(2018)Rizvi, Abbasi, and Velni}]{rizvi2018autoencoder}
Rizvi, S.Z., Abbasi, F., and Velni, J.M. (2018).
\newblock Model reduction in linear parameter-varying models using autoencoder
  neural networks.
\newblock In \emph{Proc. of the American Control Conference}, 6415--6420.

\bibitem[{Rizvi et~al.(2016)Rizvi, Mohammadpour, T{\'o}th, and
  Meskin}]{rizvi2016kernelpca}
Rizvi, S.Z., Mohammadpour, J., T{\'o}th, R., and Meskin, N. (2016).
\newblock {A kernel-based PCA approach to model reduction of linear
  parameter-varying systems}.
\newblock \emph{IEEE Transactions on Control Systems Technology}, 24(5),
  1883--1891.

\bibitem[{Rugh and Shamma(2000)}]{shamma2000gain_scheduling}
Rugh, W.J. and Shamma, J.S. (2000).
\newblock Research on gain scheduling.
\newblock \emph{Automatica}, 36(10), 1401--1425.

\bibitem[{Sadeghzadeh et~al.(2020)Sadeghzadeh, Sharif, and
  T{\'o}th}]{sadeghzadeh2020affine}
Sadeghzadeh, A., Sharif, B., and T{\'o}th, R. (2020).
\newblock Affine linear parameter-varying embedding of non-linear models with
  improved accuracy and minimal overbounding.
\newblock \emph{IET Control Theory \& Applications}, 14(20), 3363--3373.

\bibitem[{T{\'o}th(2010)}]{Toth2010}
T{\'o}th, R. (2010).
\newblock \emph{Modeling and Identification of Linear Parameter-Varying
  Systems}.
\newblock Lecture Notes in Control and Information Sciences, Vol. 403.
  Springer, Heidelberg.

\bibitem[{Varga et~al.(1998)Varga, Looye, Moormann, and
  Gr{\"a}bel}]{varga1998modelbasedred_aircraft}
Varga, A., Looye, G., Moormann, D., and Gr{\"a}bel, G. (1998).
\newblock {Automated generation of LFT-based parametric uncertainty
  descriptions from generic aircraft models}.
\newblock \emph{Mathematical and Computer Modelling of Dynamical Systems},
  4(4), 249--274.

\bibitem[{Wu et~al.(1995)Wu, Packard, and Balas}]{wu1995lpv}
Wu, F., Packard, A., and Balas, G. (1995).
\newblock {LPV control design for pitch-axis missile autopilots}.
\newblock In \emph{Proc. of the 34\tss{th} IEEE Conference on Decision and
  Control}, volume~1, 188--193.

\bibitem[{Yakimenko(2015)}]{Yakimenko2015a}
Yakimenko, O.A. (2015).
\newblock \emph{{Precision Aerial Delivery Systems: Modeling, Dynamics, and
  Control}}.
\newblock American Institute of Aeronautics and Astronautics, Inc.

\end{thebibliography}

\end{document}